\documentstyle[prl,aps,epsfig,multicol]{revtex}
\newcommand{\beq}{\begin{equation}}
\newcommand{\eeq}{\end{equation}}

\begin{document}

\title{$\frac{\lambda }{8}$-period optical potentials}
\author{B. Dubetsky and P. R. Berman}

%EndAName
\address{Michigan Center for Theoretical Physics, FOCUS Center, and Physics\\
Department, University\\
of Michigan, Ann Arbor, MI 48109-1120}

\date{\today}
\maketitle

\begin{abstract}
A Raman configuration of counterpropagating traveling wave fields, one of
which is $lin\bot lin$ polarized and the other $lin\Vert lin$ polarized, is
shown to lead to optical potentials having $\frac{\lambda }{8}$ periodicity.
Such optical potentials may be used to construct optical lattices having $%
\frac{\lambda }{8}$ periodicity. Using numerical diagonalization, we obtain
the optical potentials for $^{\text{85}}$Rb atoms.

\bigskip 
32.80.Pj, 32.80.Lg, 42.50.Vk
\end{abstract}

\begin{multicols}{2}
Recently, we proposed a new method for producing atom gratings having $%
\lambda /4$ periodicity, where $\lambda $ is the wavelength of the radiation
field driving the optical transitions \cite{c1}. The method is based on
Raman transitions that are simultaneously driven by two pairs of
counterpropagating waves. Usually, atom interactions with counterpropagating
resonant fields leads to atom gratings having overall periodicity $\lambda
/2.$ A number of techniques have been developed to reduce this periodicity,
such as harmonic suppression \cite{c2,c3,c4,c14,c41,c5}, fractional Talbot
effect \cite{c52,c15}, atom lens filtering \cite{c6,c7} and large angle beam
splitters \cite{c8,c9,c10,c11,c12,c51,c13}, but the basic starting point is
the $\frac{\lambda }{2}$ periodicity atom gratings. The scheme considered in 
\cite{c1} allows one to reduce this fundamental periodicity to $\lambda /4;$
the methods referred to above can then be used to reduced this periodicity
even further. In this brief report, we show that the basic Raman geometry
leads to optical potentials having $\lambda /8$ periodicity.

\begin{figure}
\begin{minipage}{0.99\linewidth}
\begin{center}
\epsfxsize=.95\linewidth \epsfbox{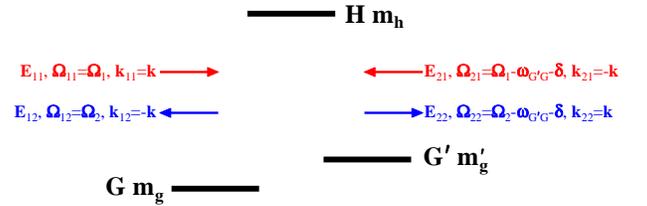}
\end{center}
\end{minipage}
\begin{minipage}{0.99\linewidth} \caption{Basic Raman configuration. 
\label{4pl}}
\end{minipage}
\end{figure}

The atom-field geometry of the Raman scheme is depicted in Fig. \ref{4pl}.
One needs {\it two} pairs of counterpropagating fields. Each pair of fields,
labeled by the subscript $j=1,2,$ itself consists of a pair of
counterpropagating fields labeled by the subscript $\alpha =1$ or $2$. The
value $\alpha =1$ corresponds to a field that drives transitions $\left|
G,m_{g}\right\rangle \leftrightarrow \left| H,m_{h}\right\rangle $ and the
value $\alpha =2$ corresponds to a field that drives transitions $\left|
G^{\prime },m_{g}^{\prime }\right\rangle \leftrightarrow $ $\left|
H,m_{h}\right\rangle $, where $H,m_{h}$ are angular momenta and Zeeman
quantum numbers of the excited state hyperfine manifold and $G,m_{g}$ and $%
G^{\prime },m_{g}^{\prime }$ are angular momenta and Zeeman quantum numbers
of two ground state hyperfine manifolds, separated by frequency $\omega
_{G^{\prime }G}$. Field ${\bf E}_{\alpha j}={\bf E}_{11}$ corresponds to a
field in the first Raman pair that drives $G\leftrightarrow H$ transitions,
field ${\bf E}_{21}$ corresponds to a field in the first Raman pair that
drives $G^{\prime }\leftrightarrow H$ transitions, field ${\bf E}_{12}$
corresponds to a field in the second Raman pair that drives $%
G\leftrightarrow H$ transitions, and field ${\bf E}_{22}$ corresponds to a
field in the second Raman pair that drives $G^{\prime }\leftrightarrow H$
transitions. Each pair of Raman fields produces two-quantum transition $%
\left| G,m_{g}\right\rangle \leftrightarrow $ $\left| G^{\prime
},m_{g}^{\prime }\right\rangle $ and the Raman detuning $\delta $ is the
same for each pair of fields$.$ However, a critical assumption of the model
is that the different pairs of Raman fields cannot interfere on
single-photon transitions, e.g. the excited state population in the $H$
manifold has no interference term associated with fields ${\bf E}_{11}$ and $%
{\bf E}_{12}$ and no interference term associated with fields ${\bf E}_{21}$
and ${\bf E}_{22}$. This can be accomplished in a number of ways - fields $%
{\bf E}_{11}$ and ${\bf E}_{12}$ can have different frequencies, different
polarizations or random frequency noise. Even though interference on single
photon transitions is suppressed, the pairs of Raman fields can interfere
and act as a {\em standing-wave Raman field }in driving transitions between
states $G$ and $G^{\prime }$. If the fields propagate along the $z$ axis and
if second Raman pair counterpropagates relative to the first, one is led to
a transition amplitude evolving as cos$\left( 2kz\right) \ $and an atomic
density having $\lambda /4$ periodicity.

Gratings involving Raman transition have been analyzed \cite
{c154,c151,c152,c153} for the case of standing waves acting on the each
optical transition. Owing to the spatially modulated ac-Stark shifts of the
atomic levels, one can achieve in this case only $\frac{\lambda }{2}$
overall periodicity. Our geometry is different. Since the fields do not
interfere on single photon transitions, the $\frac{\lambda }{2}$ periodicty
is suppressed.

In our previous article \cite{c1}, we showed that the periodicity of the
atom gratings could be reduced to $\frac{\lambda }{8}$ if one pair of Raman
fields is $lin\bot lin$ polarized and the other is $lin\Vert lin$ polarized,
e.g. 
\begin{equation}
{\bf E}_{11}\Vert {\bf E}_{12}\Vert {\bf E}_{22}\bot {\bf E}_{21}.  \label{1}
\end{equation}
This result is the Raman analogue of the conventional $lin\bot lin$
polarized field geometry for electronic transitions, which, in the
far-detuned case, leads to the $\frac{\lambda }{4}$-period atom gratings 
\cite{c16} and optical lattices \cite{c17}. The calculations of Ref. \cite
{c1} were aimed mainly at situations involving atom scattering in the
Raman-Nath approximation; however, it was pointed out in that article that
the formalism could also be applied to {\em cw} optical fields.

\begin{figure}
\begin{minipage}{0.99\linewidth}
\begin{center}
\epsfxsize=.95\linewidth \epsfbox{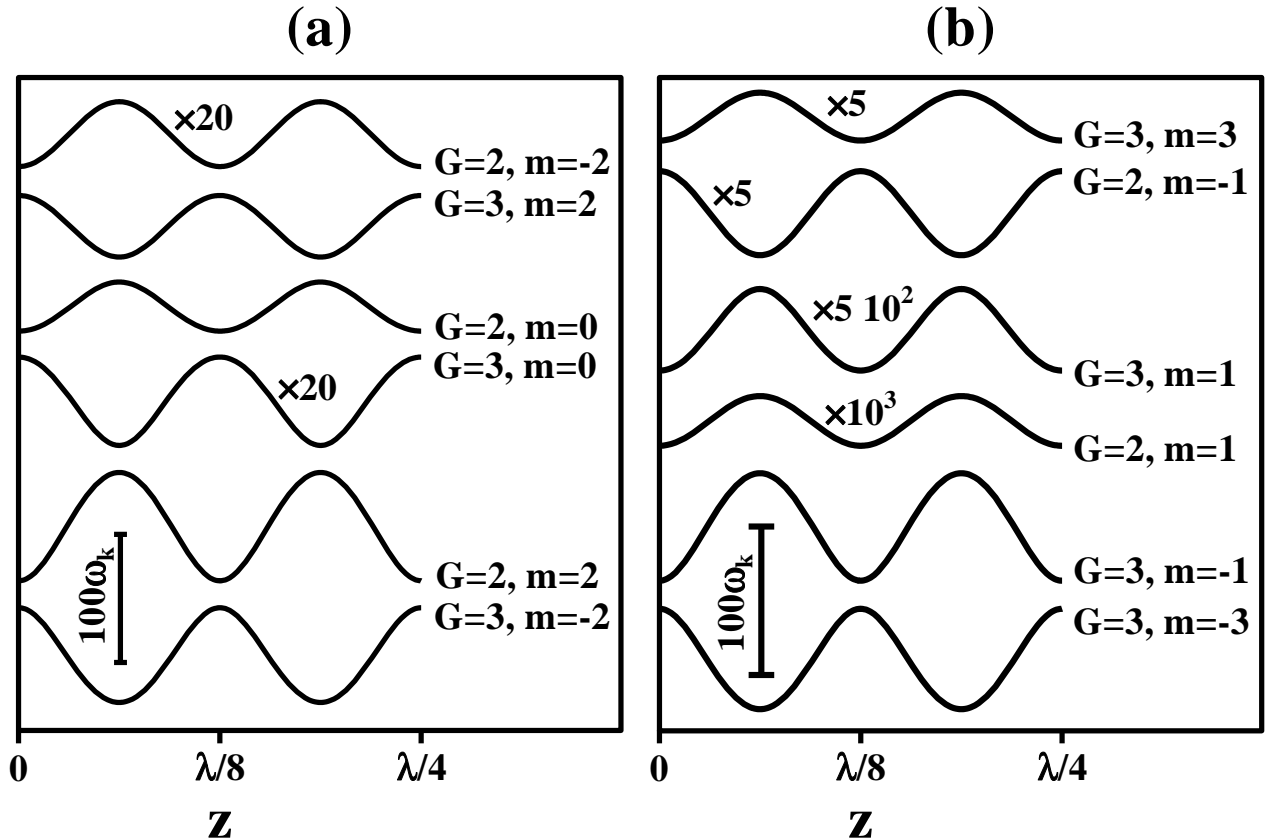}
\end{center}
\end{minipage}
\begin{minipage}{0.99\linewidth} \caption{Two groups of $\frac{\protect\lambda }{8}$-period potentials
produced on the transition between $G=2$ and $G=3$ hyperfine sublevels of $^{%
\text{85}}$Rb atoms, corresponding to the sub-systems with even (a) and odd
(b) magnetic quantum numbers. The magnetic quantum numbers correspond to the
potentials in the limit that the optical fields approach zero. 
\label{uc}}
\end{minipage}
\end{figure}

To illustrate this possibility, we proceed to calculate the optical
potentials for $^{\text{85}}$Rb when the field polarizations are given by
Eq. (\ref{1}). The resulting optical potentials have $\frac{\lambda }{8}$
periodicity and may enable one to construct optical lattices having this
periodicity. By diagonalizing numerically the Hamiltonian derived in \cite
{c1}, we obtain the optical potentials associated with the $G=2,3$ ground
state hyperfine manifolds of $^{\text{85}}$Rb. The results of the
calculations are shown in Fig. \ref{uc} and Table \ref{uct}. The optical
potentials have been displaced to fit on a single graph - mean values for
each of the potentials are listed in the Table.
\end{multicols}

\begin{table}[tbp]
\caption{Average values of the potentials shown in Fig. \ref{uc} in units of
recoil frequency $\protect\omega _{k}=\hbar k^{2}/2M.$}
\label{uct}
\end{table}

\begin{center}
\begin{tabular}{|c|c|c|c|c|c|}
\hline
$G=3,m=-2$ & $G=2,m=2$ & $G=3,m=0$ & $G=2,m=0$ & $G=3,m=2$ & $G=2,m=-2$ \\ 
\hline
-744 & -282 & -22 & 4360 & 4804 & 12781 \\ \hline
\end{tabular}

\medskip 
\begin{tabular}{|c|c|c|c|c|c|}
\hline
$G=3,m=-3$ & $G=3,m=-1$ & $G=2,m=1$ & $G=3,m=1$ & $G=2,m=-1$ & $G=3,m=3$ \\ 
\hline
-721 & 344 & -4 & -3 & 8983 & 12077 \\ \hline
\end{tabular}
\end{center}

\begin{multicols}{2}
If the quantization axis is chosen along the wave-vectors, the selection
rule for two-quantum transitions is $\Delta m_{g}=0,$ $\pm 2,$ implying that
sub-systems having even and odd Zeeman quantum numbers are decoupled from
one another, and can be diagonalized independently. We assume that all
fields drive only $D_{2}$ transitions in $^{\text{85}}$Rb, such that the
electronic angular momenta for ground and excited states are $J_{G}=1/2$ and 
$J_{H}=3/2.$ We choose field detunings $\Delta _{1j}$ for $\left|
G=2\right\rangle \leftrightarrow $ $\left| H=1\right\rangle $ transitions as 
$\Delta _{11}=2\pi \times 40$ MHz and $\Delta _{12}=2\pi \times 61$ MHz
(both detunings between the $H=2$ and $H=3$ excited state hyperfine levels)$%
, $ Poynting vectors $S_{\alpha j}=0.2\,$W/cm$^{\text{2}},$ and a Raman
detuning $\delta =2\pi \times 1.0$ MHz$.$ The eigenstates for each potential
is a $z$-dependent mixture of magnetic sublevels belonging to the different
hyperfine manifolds. Each eigenstate maps into a single magnetic substate
only when one turns off the fields. Even in this case identification of the
potentials is a problem, since magnetic sublevels for different manifolds
are degenerate. To overcome this problem we insert a small $\sim 2\pi \times
10$ KHz equidistant splitting of the sublevels. Following the smooth
dependence of the potentials' positions and amplitudes as the fields'
Poynting vector is reduced to $S_{\alpha j}=20$\thinspace $\mu $W/cm$^{\text{%
2}},$ we can assign in Fig. \ref{uc} the asymptotic identification of each
potential curve with a specific magnetic state sublevel.

It is not always possible to produce $\frac{\lambda }{8}$ period optical
potentials using the field polarizations given in Eq. (\ref{1}). In certain
limiting cases, the potentials are flat for these polarizations. For
example, if one detunes far from each hyperfine transition, fields ${\bf E}%
_{11}$ and ${\bf E}_{21}$ do not drive Raman transitions (see Eq. (11) of
Ref.\cite{c1}) and no interference between the different pairs of Raman
fields is possible. Also for transitions such as $G,G^{\prime },H=1,2,1$ or $%
G,G^{\prime },H=1,2,2$ the fact that the transition matrix elements vanish
between states having the same angular momentum and $m=0$ suppresses
interference between the different pairs of Raman fields. This implies that
the optical potentials for $^{87}Rb$ are flat with the field polarizations
given in Eq. (\ref{1}).

The possibility to produce optical potentials having a depth of a 100 recoil
energy shifts with available laser sources suggests that $\frac{\lambda }{8}$%
-period optical lattices could be constructed using the Raman technique. It
remains to calculate diffusion losses and nonadiabatic coupling to determine
the equilibrium spatial distribution of atoms in these potentials.

\acknowledgments

We are pleased to acknowledge helpful discussions with G. Raithel at the
University of Michigan. This work is supported by the U. S. Office of Army
Research under Grant No. DAAD19-00-1-0412 and the National Science
Foundation under Grant No. PHY-0098016 and the FOCUS Center Grant, and by
the Office of the Vice President for Research and the College of Literature
Science and the Arts of the University of Michigan.

\end{multicols}

\end{document}